\documentclass[article]{has40}
\usepackage{graphicx}
\usepackage{amssymb}
\tabletypesize{\normalsize}  

\def\plotone#1{\centering \leavevmode
\includegraphics[width=.95\columnwidth]{#1}}

\def\plotone#1{\centering \leavevmode
\includegraphics[width=.95\columnwidth]{#1}}

\shortauthors{Pritzl}
\shorttitle{M33 RR Lyrae Stars}

\begin{document}
\large    
\pagenumbering{arabic}
\setcounter{page}{193}

\title{Exploring M33 Through RR Lyrae Stars}

\author{{\noindent Barton J. Pritzl{$^{\rm 1}$}\\
\\
{\it (1) University of Wisconsin Oshkosh, Oshkosh, WI 54901, USA }\\
}
}

\email{(1) pritzlb@uwosh.edu}

\begin{abstract}
Recent surveys have detected RR Lyrae stars in M33, the Triangulum Galaxy. ÊThese variable stars are excellent tracers of ancient stellar populations. ÊThe RR Lyrae stars have been used to estimate metallicities at various locations within M33, as well as determining the distance to the galaxy. ÊA summary of the M33 RR Lyrae stars is presented here as well as an analysis on what their properties imply for the unique M33 galaxy.
\end{abstract}

\section{Introduction}

M33 (the Triangulum Galaxy) is a spiral galaxy similar to the Milky Way and M31 (the Andromeda Galaxy), but it also has some unique properties.  M33 shows no clear evidence of a bulge component (e.g., McLean \& Liu 1996).  McConnachie et al.\ (2009, 2010) showed that there is an extended structure surrounding M33 possibly due to tidal interaction between M33 and M31.  There has been much debate about the extent of the halo of M33.  Cockcroft et al.\ (2013) gave a good review of the past research into this topic, while using the Pan-Andromeda Archeological Survey (PAndAS) data to argue for the detection of a new component of M33 that may be the halo.

Another way of examining the properties of M33 has been through variable star surveys to search for RR Lyrae (RRL) stars.  As noted by Sarajedini (2011), the RRL stars are the ``swiss knife" of astronomy.  They can be used to determine distances, estimate metallicities and reddenings, and their detection means that the stellar population they derive from is older than about 10 billion years (a summary of RRL properties can be found in Smith 1995).  Here we summarize the RRL surveys to date and discuss their implication on the properties of M33.

\section{RR Lyrae Surveys}

The first published attempt to detect RRL stars in M33 was by Pritchet (1988).  Using the Canada-France-Hawaii 3.6 m Telescope, Pritchet, along with van den Bergh, identified RRL stars and used the seven best candidates to estimate the distance to M33.  Unfortunately, these results were preliminary, and the final results are unpublished.

Sarajedini et al.\ (2006) used observations taken by the {\it Hubble Space Telescope} (HST) to investigate two inner fields of M33 (mean deprojected distances of 13.3\arcmin).  Eight F606W and sixteen F814W images from the Advanced Camera for Surveys (ACS) were used to determine the periods of the RRL stars.  In the northwest field (U49), 29 RRab and 7 RRc stars were detected.  In the southeast field (M9) 35 RRab and 1 RRc stars were detected.  Sarajedini et al.\ (2006) used the distribution of RRab periods to determined that there were two separate populations that the RRL belonged to.  The shorter period RRL stars were from an old metal-rich disk population, while the longer period ones were from a metal-poor halo population (their metallicity being similar to M33 halo globular clusters; Sarajedini et al.\ 2000).  Within their paper, Sarajedini et al.\ (2006) acknowledge the challenge in determining periods from a limited sample is due to a lack of phase coverage.  For example, they note that some of the RRc stars have periods and amplitudes that place them in the RRab region.  Sarajedini (private communication) shared the data from Sarajedini et al.\ (2006) for a reanalysis, as discussed in Pritzl et al.\ (2011).  Pritzl et al.\ found that a number of the shorter period RRab stars had alternative longer periods or were misclassified RRc stars.  The number of RRL stars with reliable periods and their revised mean periods are listed in Table 1.  The reanalysis reduced the distinction between the shorter, disk-like and the longer, halo-like RRab stars.  Even with the reanalysis, the Sarajedini et al.\ (2006) results were important because they unequivocally proved that RRL stars can be detected in M33 and determined some basic properties about M33 from them.

\begin{flushleft}
\begin{deluxetable*}{lccccccl}
\tabletypesize{\normalsize}
\tablecaption{Properties of M33 RR Lyrae Stars}
\tablewidth{0pt}
\tablehead{ 
\colhead{Field}  & \colhead{Total RRab}  & \colhead{Total RRc}   & \colhead{$\langle P_{ab}\rangle$} &
       \colhead{$\langle P_c\rangle$} &  \colhead{$\langle [Fe/H]\rangle$}  &  \colhead{$\langle(m-M)_0\rangle$}  &  \colhead{Reference}  \\
   & & & \colhead{(days)} & \colhead{(days)} \\
}

\startdata
 U49      &  26  &    4  &  0.539  & 0.347     &  \nodata                 &  \nodata                   &  Sarajedini et al.\ (2006) \\
 M9        &  31  &    2  &  0.529  & 0.326     &  \nodata                 &  \nodata                   &  Sarajedini et al.\ (2006)  \\
 DISK2  &  65  &  18  &  0.553  &  0.325    &  $-1.48\pm0.05$  &  $24.52\pm0.11$   &  Yang et al.\ (2010)  \\
 DISK3  &  20  &    2  &  0.588  &  0.354    &  \nodata                 &  \nodata                    &  Yang et al.\ (2010)  \\
 DISK4  &  11  &    3  &  0.554  &  0.371    &  \nodata                 &  \nodata                    &  Yang et al.\ (2010)  \\
 SE45   &     1  &    1  &  0.630  &  0.428    &  \nodata                  &  \nodata                    &  Pritzl et al.\ (2011) \\
 SE25   &   12  &   0   &  0.628  & \nodata  & $-1.84\pm0.30$   &  $24.69\pm0.17$   &  Pritzl et al.\ (2011) \\
 \enddata
\end{deluxetable*}
\end{flushleft}

In a follow-up to Sarajedini et al.\ (2006), Yang et al.\ (2010) investigated three fields to the southwest of M33 using {\it HST/ACS}.  These observations consisted of twenty F606W and twenty-four F814W images.  The fields have galactocentric distances of 9.3\arcmin (DISK2), 15.9\arcmin (DISK3), and 22.6\arcmin (DISK4).  A total of 119 RRL variables were found (see Table 1).  Their conclusion from the properties of the RRL stars is that these do belong to the halo.  Their mean metallicity was determined to match well with the M33 halo globular clusters.  

Using the GMOS-N instrument at the Gemini North Observatory, Pritzl et al.\ (2011) detected RRL stars within two southeast fields.  Sixteen $g'$ images were used to determine the periods of the RRL stars, while four $r'$ images were used to help determine colors of the stars.  The outermost field (SE45) was $45\arcmin$ from the center of M33, which leads to a deprojected distance of about 19 kpc.  The second field (SE25) was $25\arcmin$ from the center of M33 for a deprojected distance of about 10 kpc.  The mean properties of the RRL stars are listed in Table 1.

\begin{figure*}
\centering
\plotone{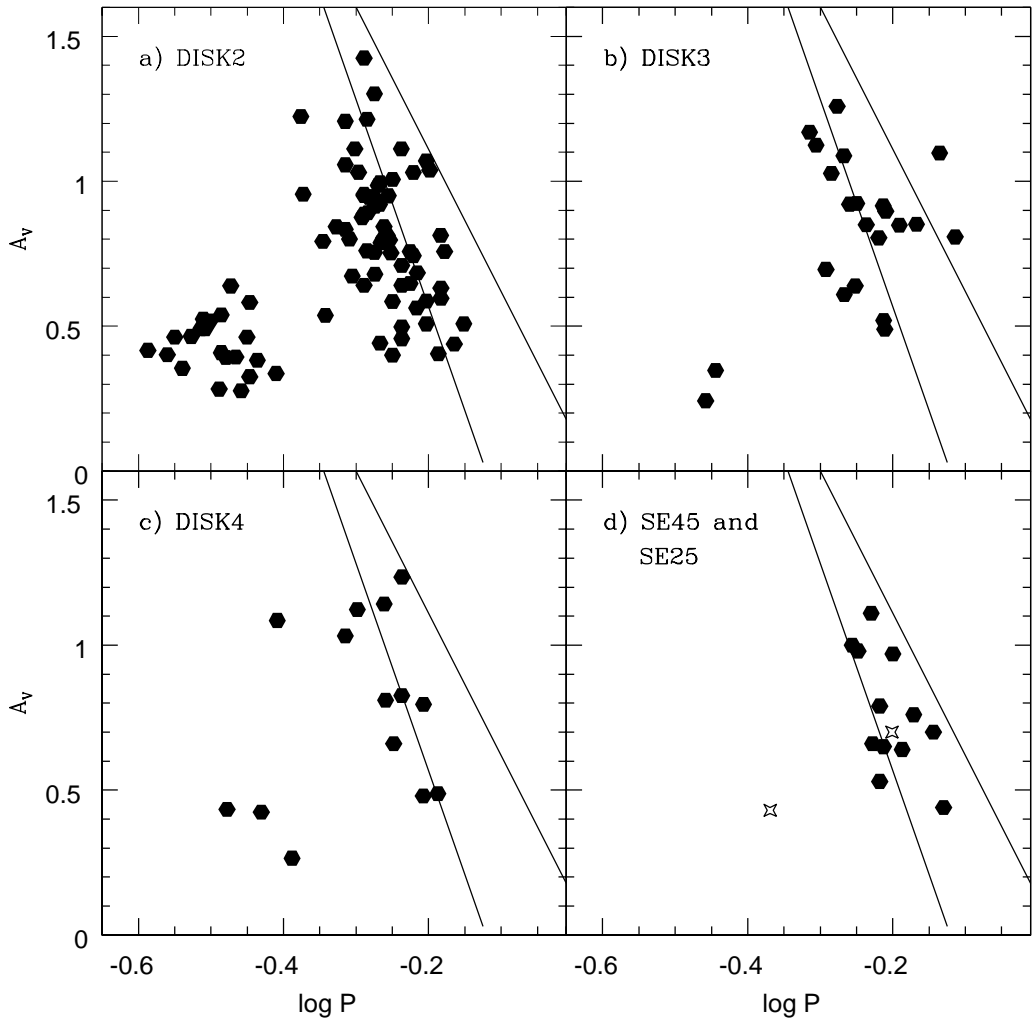}
\vskip0pt
\caption{Period-amplitude diagrams for RR Lyrae stars in M33.  Figures a-c are the three fields from Yang et al.\ (2010) starting from the innermost field.  Figure d is the fields from Pritzl et al.\ (2011).  The filled points are from the SE25 field, while the open stars are from the SE45 field.  The lines represent the Oosterhoff I and II lines taken from Clement \& Rowe (2000).}
\label{fig1}
\end{figure*}

\section{Analysis and Summary}

As noted in the introduction, RRL stars are useful for determining distances.  As seen in Table 1, the distance moduli determined from two of the variable star fields are within the margin of error.  In addition, Sarajedini et al.\ (2006) found a distance modulus of $24.67\pm0.08$.  These measurements are consistent with two measurements using the tip of the red giant branch method ($24.64\pm0.08$, Galleti et al.\ 2004; $24.69\pm0.07$, Tiede et al.\ 2004).  

In comparing the properties of the RRL stars, we can see some interesting trends.  The data from the inner fields (Sarajedini et al.\ 2006) and the southwest fields (Yang et al.\ 2010) point to an RRL population that is Oosterhoff I (relatively metal-rich, shorter period RRL stars).  Although smaller in number, the RRL stars in the outer southeast fields (Pritzl et al.\ 2011) are more in line with the Oosterhoff II classification (metal-poor, longer period RRL stars).  This can also be seen in the period-amplitude diagrams seen in Figure 1.   The inner and southwest fields are closer in metallicity to the halo globular clusters ([Fe/H]$=-1.27\pm0.11$; Sarajedini et al.\ 2000) and field stars ([Fe/H]$=-1.24\pm0.04$; Brooks et al.\ 2004), as noted by Yang et al., whereas the outer southeast fields are much more metal-poor.  It should be noted that the metallicities for the RRL stars were determined using a formula determined by Sarajedini et al.\ (2006) using pulsation properties of Galactic RRL stars.

This raises the question as to which populations belong to the halo, if any?  Could the longer period RRL stars in Pritzl et al.\ (2011) be from the halo, while the shorter period RRL stars in Sarajedini et al.\ (2006) and Yang et al.\ (2010) are from the disk?  It should be noted that the M9 field in Sarajedini et al.\ is nearly in-line with the fields in Pritzl et al.\ extending toward the southeast.  Perhaps there is some kind of radial gradient in the stellar population in that direction?  Clearly more research needs to be done on the variable star population in M33 to answer these questions.

\acknowledgements
Thank you to A. Sarajedini for discussions on M33 and for sharing data.  A special thank you to Horace A. Smith for his guidance and support over all of these years!

\end{document}